# Cyberbullying Indicator as a Precursor to a Cyber Construct Development

**Salam Khalifa Al-Romaihi[1] and Richard Adeyemi Ikuesan[2]**
**[1]Community College Qatar, Doha, Qatar**
**[2]Zayed University, Abu Dhabi, UAE**
Salman63962@ccq.edu.qa
Richard.Ikuesan@zu.ac.ae

**Abstract**: The current global pandemic occasioned by the SARS-CoV-2 virus has been attributed, partially, to the growing range of cyber vises within the cyber ecosystem. One area of such impact is the increasing tendencies of cyber-bullying among students. Cyberbullying -the act of subjugating others using a cyber platform- is a growing concern among educators, especially in High-Schools. Whilst studies have been carried out towards understanding this menace, the approach towards identifying indicators of cyberbullying is largely missing in the literature. To address this research gap, this study proposed a cyberbullying framework based on the identification of some observable behavioral indicators. Using a self-administered measurement instrument from 30-respondents, the study observed the probability of a cyberbully construct, as a potential measure of the presence of cyberbullying; a probability that has been largely ignored in extant literature. This observation presents a veritable tool for the development of an active and integrated learning platform void of abuse among students. Furthermore, within the cyber education ecosystem, a cyberbullying construct would provide a mechanism for the development of an appropriate online learning platform, which would be useful to the information system and cyber education research communities.

## 1. Introduction

Like physical bullying, cyberbullying can occur in diverse ways, and can also reflect in myriad forms, ranging from direct to indirect, intentional to unintentional, one-to-one, one-to-many, and many-to-many as well as overt-to-covert actions. However, unlike physical bullying, cyberbullying can transcend borders, boundaries, and even temporal-spatiality with a long-lasting effect. For instance, any act posted online could be retained for years until such an act is repealed or taken down. The concept of cyberbullying, as highlighted in Corliss, (2017) and Lee and Chun, (2020), is rooted in diverse ontological definitions which primarily involve the use of technology. Put succinctly, cyberbullying is the act of leveraging the power of technology to cause harm or discomfort to another. This act is often associated with frequent or repeated communication occasioned by aggressive tendencies (towards a hurtful, disgraceful, harmful, or discomfort outcome) to recipients (Nakano *et al.*, 2016; Buelga *et al.*, 2020). Redmond, Lock and Smart, (2020) further dissected technology used to include electronic mails, telecommunication text messages, chat rooms, mobile phones, web, and mobile cameras, as well as websites. With the steady growth and adoption of technology, the cyberbullying act can be projected to increase. Different dimensions of cyberbullying acts have been identified in the literature (Langos, 2012). Cyber personality has also been defined in existing studies (Adeyemi *et al.*, 2016; Adeyemi, Abd Razak, *et al.*, 2017; Adeyemi, Razak and Salleh, 2017; Adeyemi, Razak, *et al.*, 2017; Ikuesan *et al.*, 2019) with contradictory cyber:physical personalities (Gardini, Cloninger and Venneri, 2009; Mohammadi and Vinciarelli, 2015; Ikuesan, Abd Razak and Salleh, 2016). The next section explores the different dimensions of cyberbullying with a view to highlighting the underlying commonalities.

## 2. Cyberbullying Dimensions

In trying to understand the different dimensions of cyberbullying, it is essential to fully understand what cyberbullying is. Within the research community, cyberbullying has garnered diverse interpretations and consequently, different meanings (Langos, 2012; Redmond, Lock and Smart, 2020). Building on traditional bullying, as measured by the Olwens Bullying questionnaire, studies have revealed that respondents often equate cyberbullying to acts that involve the internet and technology (Dennehy *et al.*, 2020; Fernández-Antelo, Cuadrado-Gordillo and Parrà, 2020). For example, the item "somebody used the internet or a cellphone to harm/offend me" has been regarded as an act of cyberbullying. As elaborated in existing studies (Corliss, 2017; Redmond, Lock and Smart, 2020), these definitions and perspectives, as a construct, present a complex phenomenon that requires more insight. Position in Langos, (2012) summarized the existing definitions within the context of intention: direct or indirect. The study coined the definition as "*Cyberbullying involves the use of ICTs to carry out a series of acts as in the case of direct cyberbullying, or an act as in the case of indirect* cyberbullying, intended to harm another (the victim) who cannot easily defend him or herself". The definition reveals cyberbullying to include acts of willfulness, power imbalance, repetitiveness, perception-based cognition, as well as a technology-integrated process. Moreover, the integration of "cyber" into traditional bullying presents a susceptibility and vulnerability beyond temporal-spatial boundaries, which further enhanced the proclivity for bullying. Therefore, irrespective of the perspective of the definition from diverse stakeholders, the underlying tenet of cyberbullying remains visible for evaluation. Evaluating cyberbullying, its effect, and potential mitigation approaches can be hinged the process of identifying cyberbullying cues, hereinafter refers to indicators. A study in (Redmond, Lock and Smart, 2020) presented three perspectives of cyberbullying indicators based on bully attributes, using a taxonomical structure. The study leveraged bully attributes, bullying categories, awareness based on the victim's and educator's perspective. These attributes, further classed as Category-1, provide a baseline for cyberbullying identification within the cyberbullying conceptual framework. Whilst the identification of cyberbullying act, specifically from the student's perspective, presents a means for cyberbullying construct development (Dennehy *et al.*, 2020; Fernández-Antelo, Cuadrado-Gordillo and Parrà, 2020), the content presented in the conceptual framework lacks useful composition that can be used as indicators or precursors. Cyber bully indicator can be defined within a context or in perspective: the victim, perpetrator, or collaborators, for example. These perspectives are further distilled using the modified Olwens' bullying cycle shown in Figure 1. Two potentially applicable perspectives (E2 and E3) are integrated into the cycle: unwilling supporter and unconscious supporter. These perspectives has been clustered together in some studies to mean unintentional bullies (Langos, 2012). However, recent advances in technology have provided a baseline for the separating these two perspectives.

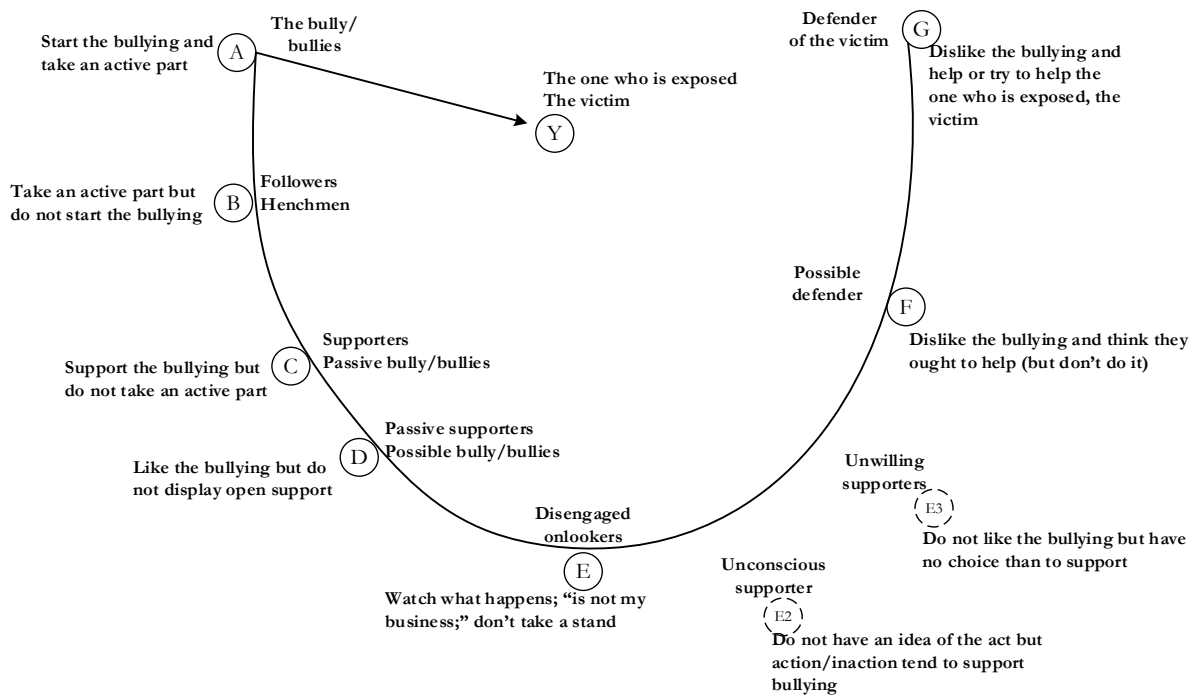

**Figure 1**: Modified Olwen's bullying cycle.

An unwilling supporter is an entity who does not like bullying but thinks he/she has no option other than to partake. This entity can also be termed a victim of the bullying act. The second category; an unconscious supporter, is an entity whose action/inaction unknowing encourages bullying. This category could potentially form a large population. This is particularly applicable to a cyber-induced bullying process where entities unknowingly provide support to abusive acts that could harm others. Therefore, approaches that provide measures for reliably identifying signs and markers of cyberbullying are largely missing in the literature. Indicators in this regard refer to a measurable instrument that can be used to deduce the tendency and or presence of cyberbullying act/process. To address this observed limitation, this study explores the probability of extracting inherent markers of cyberbullying. The study further proposed an identification mechanism rooted in the result of the exploratory process. To the best of the Authors' knowledge, this study serves as a pioneering insight towards the development of a reliable construct, hereinafter referred to as the Cyber Bullying Indicator construct (CBIC). Within the general body of information systems, a construct is a concept of abstraction that cannot be directly described using a singular entity, however, exist, nonetheless. Constructs such as Perceived

Ease of Use (PEU), Perceived Usefulness (PU), and Perceived Behavioral Control (PBC) are examples of constructs that have been widely used in the literature. This study, therefore, presents a foundational basis on which a CBIC can be developed. This is achieved based on an underlying research question given thus: *Given the growing tendency of online cyberbullying in educational settings, can the use of a quantitative exploratory study be used to identify any underlying constructs that can explain cyberbullying with a Cronbach's alpha reliability greater than 0.7?*

The remainder of the manuscript is organized as follows: the next section discusses the methodology applied in this study to explore the conceptualization of the construct. This is followed by the presentation of the exploratory result and an in-depth analysis of the result. Conclusion and future works are thereafter provided.

## 3. Methodology

Given the nature of cyberbullying, and the scope of this study, an exploratory quantitative approach was considered. A set of measurement instruments was developed based on the exploration of common cyberbullying concepts used in literature. The final compilation comprised 20-items. Each item represents a potential measure of bullying indicator. This includes the two ends of the bullying cycle spectrum identified in Figure 1: the bully (A) and the victim (Y). To select participants for this study, three criteria were defined as shown in Table 1. Taken together, the criteria provide a guide for respondent selection.

**Table 1**: Respondent Inclusion Criteria

| Criteria | Description |
|---|---|
| 1 | Respondents must be in academics in departments that work with information technology either for duties or develop information technology content. |
| 2 | Respondents should have knowledge of cyberbullying, either from a behavioral science or experience. This can be within the social science discipline. |
| 3 | Respondents must be a parent who has a child(ren) in secondary school(s) |

Upon receipt of ethical clearance from the institution, the preliminary exploratory process was conducted. The questionnaire instrument was sent to selected Faculties in the institution via email. This process leveraged the purposive non-probability sampling technique as opposed to the probabilistic random sampling technique. The summary of the measurement instrument developed for this study is further shown in Table 2. The instrument contains 20 items adapted from existing studies, which can be further grouped into two classes.

**Table 2**: Measurement Instrument for CBIC

| Item | Content | Factor |
|---|---|---|
| Q3 | Abruptly shutting down or walking away from the computer mid-use | Phase-1 |
| Q6 | Expresses abrupt reaction when his/her online-related gadget is being touched | |
| Q9 | Upset or frustrated after going online or gaming | |
| Q12 | Withdrawing from friends and family in real life after a long hour of techno-isolation | |
| Q14 | He or she clears the computing device screen abruptly when someone enters the room. | |
| Q16 | He/she starts becoming secretive about their Internet activity. | |
| Q18 | Does not like to use websites they used to spend time on. | |
| Q19 | He/she seems upset, highly irritable, or emotional after being on the computer | |
| Q5 | Unexplained anger or depression, especially after going online | Phase-2 |
| Q8 | Sudden detachment from electronic appliance | |
| Q11 | Demonstrates uneasy, nervous, or scared when he/she receives an electronic message | |
| Q17 | There is a sudden drop in computer use and seems to avoid its use | |
| Q20 | He/she seems upset, highly irritable, or emotional after reading their text messages or email on their mobile device | |

A total of 30 responses were received. Whilst the sample size of 30-response is relatively small, the authors assert that the sample presents a baseline for further studies. As highlighted in Lee and Chun (2020), studies on conceptualizing cyberbullying can leverage a sample size greater than 10-response. The sample in the current study is relatively higher than the sample used in Phase-2 of the concept mapping process in Lee and Chun (2020). Given that this is the preliminary insight into the subject matter, the sample size was considered sufficient for this study. Statistical Package for Social Science (SPSS) v23 was used to conduct the exploratory process while SMART PLS (student License) was used to explore confirmatory factor analysis. The detail of the result is presented in the next section.

## 4. Result and Analysis

An exploratory analysis typically involves the process of extracting the underlying intrinsic cluster in data. To do this, the Cronbach's Alpha test of instrument reliability was carried out. The result showed that the items were indeed reliable with a Cronbach's Alpha value of 0.803. The Kaiser, Meyer, Olkin (KMO) measure of sampling adequacy generated poor sphericity at 0.346 which was an expected outcome, given the overall sample size used in this study. In the initial factor exploration, no pre-defined number of clusters was considered (which implied an Eigenvalue > 1 was considered). A total of 5-underlying factors was extracted using maximum likelihood as the method of extraction. A further observation of the result showed that the data has two underlying factors. Thus, pruning the factor into two and suppressing coefficients< 2, the pattern analysis revealed two overlapping factors, at a $\chi^2/df = 1.612$. From the extracted pattern, 4-items had relatively similar cross-loading. These items were deleted. Furthermore, 7-items and 9-items were classified as factors 1 and 2 respectively. To perform the confirmatory analysis, the pruned data was used to model the path analysis. The path model revealed the factor loading of three items to be less than 0.4. These items were further removed. The resultant path model is presented in Table 3 and Figure 2. The observed path coefficient between the two factors was 0.55.

**Table 3**: Construct Reliability and Validity

| Factor Description | Cronbach' Alpha | Rho Alpha | Composite Reliability (CR) | Average Variance Extracted (AVE) | R Square |
|---|---|---|---|---|---|
| Phase-1 | 0.831 | 0.842 | 0.87 | 0.463 | |
| Phase-2 | 0.712 | 0.749 | 0.739 | 0.446 | 0.308 |

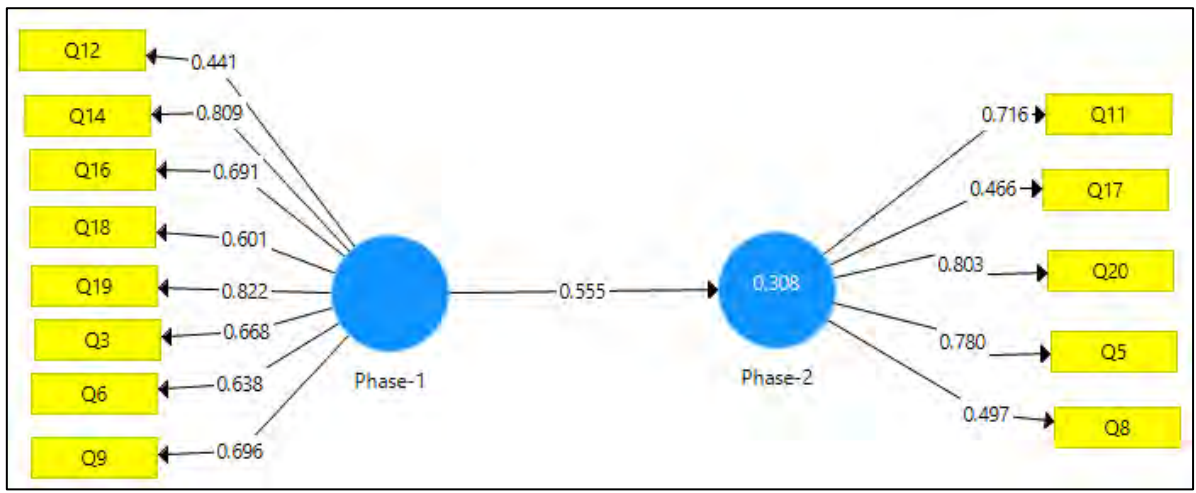

**Figure 2:** Path Model towards a confirmatory factor analysis

From the result shown in Table 3, it certainly appears that the two factors can model the measurement items. The factor loading for all items in Factor-1 (hereinafter referred to as Phase-1) and Factor-2 (hereinafter referred to as Phase-2) are greater than 0.4. In terms of construct reliability and validity, both factors satisfied the standard thumb rule for the Cronbach's alpha, rho alpha, and composite reliability. However, the AVE was below the 0.5 thumb rule. Using Phase-1 as a predictor variable for Phase-2 and vice versa, the study generated an identical regression weight (0.555) which further shows the probable existence of two variables that can be used to measure the proposed cyberbullying construct.

## 5. Discussion

The result presented in the previous section revealed the probability of the existence of a cyberbullying indicator construct. A cyberbullying construct provides a baseline for measuring the presence of bullying, which can be measured by parents, guardians, as well as educators. Often, subtle bullying process among secondary school students can go undetected until it causes damages to the victims. This study, therefore, serves as a first step towards mitigating cyberbullying act, especially among secondary school students. With such an indicator, different stakeholders can identify signs before it degenerates into psychological and physical trauma. This is especially useful given the trendy adoption of digital lifestyle occasioned by the Covid-19 pandemic and the potential for an inseparable work-life balance for parents. Cyberbullying is relatively difficult to identify without the core indicators, relative to physical bullying (Lee and Chun, 2020). Indicators in this regard reveal necessary precursors that can be used to glean the emergence or inception of a cyber induced bullying process. A summary of the final measurement is given in Table 2. A breakdown of the item's clusters in Table 2 shows that there exist some underlying tendencies that can be further classified into Phase-1 and Phase-2 behavior. Both Phases 1 and 2 depict a reflective indicator that shows the behavior of an entity after an incident has occurred. For example, the concept of abrupt change in behavior after an online experience could reflect what had transpired during an Internet surfing period. This behavior could result in excitement on the part of the bully, while the converse would be seen on the victim. The items defined within this study can be further examined in line with the underlying tenet in each item. Though words like withdrawer, abrupt, highly irritable, emotional, as well as upset, are general terms associated with victims of a bully, the concept of cyber content is revealed in words like an electronic message, internet gaming, online activities, techno-isolation, computing device, and online-gadget. These keywords can be associated with cyber activities. Though the current exploratory analysis generated two clusters, a confirmation of this would be rigorously carried in the subsequent study. This is essentially required to formulate a steady construct for the CBIC. A theoretical analysis of the measurement items in each Phase did not reveal a remarkable distinction. This further explains the need for further study.

Ideally, a construct consists of a singular class, and the process to formally develop such would pass through a series of reliability tests. Further study with a significantly larger sample size will be carried out to ascertain the feasibility of developing the CBIC. As an ongoing work, a further study that involves the development of a nomological network, as well as external factors that can be used to validate the probability of the existence of the CBIC, will be carried out. This attempt will leverage constructs from other information systems models such as the theory of planned behavior and the extended technology acceptance model. Also, other cyberbullying constructs such as the CYB-AGS (Buelga *et al.*, 2020) aggressive scale can be explored. Such a network can be used to measure the impact of cyber education and awareness against potential bullies. One aspect that is not considered in this study is the influence of the internet as a tool that aids anonymity. This notion can have a potential impact on who becomes a bully, and the transformative tendencies of cyberbullying. For instance, some studies (Haddadain, Abedin and Monirpoor, 2010; Yue *et al.*, 2010; Sari, 2016) have shown that the anonymity provided by the Internet induces different kinds of personalities on the Internet. This could further influence the degree of power imbalance associated with bullying. Opinion in Dennehy *et al.*, (2020) argued that the bully's "inability to witness the victim's reaction may diminish their empathetic response potentially leading to more harshness in the cyber behaviors". In addition, online anonymity could induce unintentional bullying, which existing studies have not explored. Online anonymity can be perceived differently between the bully and the victim. Further study on the reaction of Victims to cyberbullying is an aspect that can provide concrete insight into this perspective. Conjecturing or attempting to hypothesize on the possible behaviorism of the victim without empirical proof, could be misleading.

## 6.  Conclusion

The concept of cyberbullying has been transformed with the advances in technology. This study explored the potential of developing a construct that can be used by different stakeholders to pinpoint and or identify the inception of cyberbullying act. Using an exploratory process, the study observed with high reliability, the probability of the existence of such a construct, which is further defined as cyberbullying indication construct (CBIC). However, the sample size used in this current study presents a major limitation of the study. A CIBC, when fully explored on larger sample size, can be a veritable tool for mitigating cyberbullying among vulnerable entities. Such a tool can be leveraged by the wider scientific community in advancing cyberbullying research. Furthermore, the integration of the proposed construct with other information systems constructs can be used to develop a nomological network for studying the effect of cyberbullying on students, as well as workplace culture.